\documentclass[11pt]{article}
\usepackage{amsmath}
\usepackage{amsfonts}
\usepackage{amssymb}
\usepackage{theorem}
\newtheorem{theorem}{Proposition}
\newcommand{\dbar}{\bar{\partial}}
\newcommand{\wt}{\widetilde}
\newcommand{\be}{\begin{equation}}
\newcommand{\ee}{\end{equation}}
\newcommand{\bea}{\begin{eqnarray}}
\newcommand{\eea}{\end{eqnarray}}
\newcommand{\beaa}{\begin{eqnarray*}}
\newcommand{\eeaa}{\end{eqnarray*}}

\newcommand{\nn}{\nonumber}
\begin{document}
\title
{Nonlinear Beltrami equation and $\tau$-function 
for dispersionless hierarchies}
\author{
L.V. Bogdanov\thanks
{L.D. Landau ITP, Kosygin str. 2,
Moscow 119334, Russia} 
~and B.G. Konopelchenko\thanks
{Dipartimento di Fisica dell' Universit\`a di Lecce
and Sezione INFN, 73100 Lecce, Italy}}
\date{}
\maketitle
\begin{abstract}
It is proved that the action for nonlinear Beltrami equation
(quasiclassical $\dbar$-problem) evaluated on its solution gives
a $\tau$-function for dispersionless KP hierarchy. Infinitesimal transformations
of $\tau$-function corresponding to variations of $\dbar$-data are found.
Determinant equations for the function generating these transformations are derived.
They represent a dispersionless  analogue of singular manifold (Schwar\-zi\-an) 
KP equations. Dispersionless 2DTL hierarchy is also considered.
\end{abstract}
\section{Introduction}
Dispersionless integrable hierarchies attracted a considerable
interest during the last ten years (see e.g. \cite{KM77}-\cite{13}).
Recently it became clear that they play an important role
in various problems of hydrodynamics and complex analysis
\cite{14}-\cite{DMT}.

Dispersionless integrable hierarchies can be described in different
forms within different approaches. In the papers \cite{KMR,KM1,KM2}
it was shown that such hierarchies can be introduced starting with the 
nonlinear Beltrami equation (quasi-classical $\dbar$-problem)
\bea
S_{\bar z}=W(z, \bar z, S_z),
\label{ddbar}
\eea
where $z\in\mathbb{C}$, bar means complex conjugation,
$
S_z=\frac{\partial S(z,\bar{z})}{\partial z},
$
$
S_{\bar z}=\frac{\partial S(z,\bar{z})}{\partial \bar z},
$
and $W$ (quasi-classical $\dbar$-data) is an analytic function
of $S_z$. Applying the quasi-classical $\dbar$-dressing method
based on equation (\ref{ddbar}), one can get dispersionless
integrable hierarchies and the corresponding addition formulae
in a very regular and simple way. Such an approach reveals also
the connection of dispersionless hierarchies with the quasi-conformal
mappings on the plane.

In the present paper we demonstrate that the quasi-classical
$\dbar$-dressing method based on equation (\ref{ddbar})
leads to explicit formula for the $\tau$-function for dispersionless
hierarchies, which is connected with 
the Lagrangian for equation (\ref{ddbar})
and the corresponding action, evaluated on the
solution of the boundary problem for this equation. We will also
derive determinant form of the generating equations for the function
$S_z$, defining infinitesimal deformations of the $\tau$-function.
In this paper we concentrate on the dispersionless 
Kadomtsev-Petviashvili (dKP) hierarchy, but we present also the basic 
formulae for the dispersionless 2DTL hierarchy.
\section{$\tau$-function as an action for nonlinear
Beltrami equation}
It was shown in \cite{KMR,KM1} that the dKP hierarchy
is connected with Beltrami equation (\ref{ddbar})
with the $\dbar$-data equal to zero outside the unit disc.
This problem can be formulated as a boundary problem
for equation (\ref{ddbar}) in the unit disc as follows.
Let the function $S_0(z)$ analytic in the unit disc $D$ be given.
The problem is to find the function $S=S_0+\wt S$, satisfying
(\ref{ddbar}), with $\wt S$ analytic outside the unit disc
and decreasing at infinity (this is in fact a boundary condition
on the unit circle, which can be written down 
using standard projection operator). We suggest that the function
$W$ is of the form 
\bea
&&
W(z,\bar z,S_z)=\sum_{p=0}^{\infty}w_p(z,\bar z)(S_z)^p,
\label{Wform}
\eea
where $w_p(z,\bar z)$ are arbitrary smooth functions in the unit disc
vanishing on the boundary.

Introducing parameterization of the function $S_0(z)$ in 
terms of times, $S_0(z)=\sum_{n=1}^\infty t_n z^n$, and using
the technique of quasi-classical $\dbar$-dressing method, 
it is possible to demonstrate that $S(z,\mathbf{t})$ is a solution
of Hamilton-Jacobi equations
for dKP hierarchy, and the first coefficient of expansion
of $\wt S(z,\mathbf{t})$ as $z\rightarrow\infty$ satisfies
equations of dKP hierarchy (see \cite{KMR,KM1,KM2,BKA}).

Here we establish a relation between the action for the 
problem (\ref{ddbar}) and the $\tau$-function for dKP hierarchy.
This relation illustrates a well-known observation that a transition
from dispersionfull to dispersionless hierarchies resembles
a transition from quantum mechanics to classical mechanics.

It was noted in \cite{KM2} that equation (\ref{ddbar})
is a Lagrangian one. It can be obtained by variation of the action
(for the boundary problem in the unit disc)
\bea
f
=-\frac{1}{2\pi\text{i}}\iint_{D} 
\left(\frac{1}{2}\wt S_{\bar z} \wt S_z -
W_{-1}(z,\bar z,S_z)\right)dz\wedge d\bar z,
\label{action}
\eea
where
\bea
W_{-1}(z,\bar z,S_z)=\sum_{p=0}^{\infty}w_p(z,\bar z)
\frac{(S_z)^{p+1}}{p+1},
\qquad
\partial_\eta W_{-1}(z,\bar z,\eta)=W(z,\bar z,\eta).
\nn
\eea
One should consider independent variations of $\wt S$,
possessing required analytic properties (analytic outside
the unit circle, decreasing at infinity),
keeping $S_0$ fixed.
\begin{theorem}
The function
\bea
F(\mathbf{t})
=-\frac{1}{2\pi\text{i}}\iint_{D} 
\left(\frac{1}{2}\wt S_{\bar z}(\mathbf{t}) \wt S_z(\mathbf{t})-
W_{-1}(z,\bar z,S_z(\mathbf{t}))\right)dz\wedge d\bar z,
\label{TAU}
\eea
i.e., the action (\ref{action}) evaluated on the solution 
of the problem (\ref{ddbar}), is a $\tau$-function of 
dKP hierarchy.
\end{theorem}
\textbf{Proof}
In order to prove that $F(\mathbf{t})$ is a $\tau$-function
of dKP hierarchy, it is sufficient to demonstrate that 
(see, e.g., \cite{BKA})
$$
\wt S(z,\mathbf{t})=-D(z)F(\mathbf{t}),
$$
where $D(z)$ is the quasiclassical vertex operator,
$D(z)=\sum_{n=1}^{\infty}\frac{1}{n}\frac{1}{z^n}
\frac{\partial}{\partial t_n}$, $|z| > 1$.
Applying the operator $D(z)$ to the r.h.s. of formula 
(\ref{TAU}), one gets (we change the variable of integration
to $y$)
\bea
D(z)W_{-1}(y,\bar y,S_y(\mathbf{t}))=
\wt S_{\bar y}D(z)(\wt S_y+ {S_0}_y).
\nn
\eea
Using the formula
$$
D(z){S_0}_y=\frac{1}{z-y},
$$
we get
\bea
\frac{1}{2\pi\text{i}}\iint_{D} 
\wt S_{\bar y} D(z){S_0}_y dy\wedge d\bar y=-
\frac{1}{2\pi\text{i}}\oint\frac{1}{z-y}\wt S(y) dy=-
\wt S(z).
\nn
\eea
Thus we have obtained the crucial term for our proof,
and now we should demonstrate that the combination of all other
terms, namely, 
\bea 
-\frac{1}{4\pi\text{i}}\iint_{D} 
(\wt S_{y}D(z)\wt S_{\bar y}-\wt S_{\bar y}D(z)\wt S_y)dy
\wedge d\bar y
\nn
\eea
is equal to zero. Indeed, using Green's formula and taking
into account that $\wt S_{\bar y}$ is equal to zero on the unit circle,
and the function $\wt S(y)$ is analytic outside the unit circle
and decreases at infinity,
we immediately come to the conclusion that this combination
is equal to zero. QED
\\
\textbf{Remark 1.} Note that integral formulae for the $\tau$-function
of different type has been derived also in \cite{8,WZ}.\\
\textbf{Remark 2.} Let us consider also the integral
\bea
F_s(\mathbf{t})
=-\frac{1}{2\pi\text{i}}\iint_{D} 
\left(\frac{1}{2} S_{\bar z}(\mathbf{t})  S_z(\mathbf{t})-
W_{-1}(z,\bar z,S_z(\mathbf{t}))\right)dz\wedge d\bar z.
\label{TAUs}
\eea
Since $\dbar S_0=0$ for $z\in D$, using Green's formula, one
obtains
$$
F_s=F+\frac{1}{4\pi\text{i}}\oint dz \wt S(z) {S_0}_z.
$$
Taking into account the relation $\wt S(z)=-D(z)F$, one gets
\bea
F_s=F-\frac{1}{2}\sum_{n=1}^\infty t_n
\frac{\partial F}{\partial t_n}.
\label{Fs}
\eea
The dKP hierarchy and addition formula for $F$ admits scale
invariance 
$$
F(\mathbf{t})\rightarrow F'(\lambda\mathbf{t})=
\lambda^2F(\mathbf{t}).
$$
A full infinitesimal variation of $F$ under this transformation
is
$$
\delta_s F=\delta_\text{form} F + \delta_\text{t} F=
-2\epsilon F,
$$
where $\delta_\text{form} F$ denotes a variation of the form of
$F$, while $\delta_\text{t} F$ stands for the variation due to the
infinitesimal variation of times $\mathbf{t}$.
In virtue of (\ref{Fs}), one has
$
\delta_\text{form} F=2\epsilon F_s.
$
In the particular case $W(z, \bar z, S_z)=\mu(z,\bar z) S_z$ we have
$F_s=0$ and thus $F$ is a homogeneous function of times of the
second order. Such a class of $\tau$-functions has been considered
within different approaches in \cite{14,TT}. Starting with linear 
Beltrami equation, we obtain only a subclass of $\tau$-functions,
which are quadraic with respect to times.\\

It is well known that dKP $\tau$-function
$F$ obeys 
the dispersionless addition formula \cite{TT,TT1}
\bea
(z_1-z_2)e^{D(z_1)D(z_2)F}+(z_2-z_3)e^{D(z_2)D(z_3)F}
+(z_3-z_1)e^{D(z_3)D(z_1)F}
=0,
\label{dKPadd}
\\
z_1,z_2,z_3\in\mathbb{C}\setminus D.
\nn
\eea 
The formula (\ref{TAU}) gives a solution to this equation
in terms of solution of nonlinear Beltrami equation (\ref{ddbar}).
\section{Variations of the $\tau$-function}
The function $W$ is the $\dbar$ data for the dKP hierarchy.
Its variations provide us with a wide class of variations of the
function $F$. For the functions $W$ of the form (\ref{Wform}),
varying $w_n(z,\bar z)$,
one has 
\beaa
\delta W=\sum_{n=1}^{\infty}\delta w_n(z,\bar z)\left(S_z\right)^n,
\qquad
\delta W_{-1}=\sum_{n=1}^{\infty}
\frac{\delta w_n}{n+1}\left(S_z\right)^{n+1},
\eeaa
and 
\bea
\delta F
=\frac{1}{2\pi\text{i}}
\iint_{D} (\delta W_{-1})(z,\bar z,S_z)dz\wedge d\bar z.
\label{Fvar}
\eea
Considering elementary variation 
$\delta w_{n_0}=\epsilon \alpha_{n_0} \delta(z-z_0)$,
$\delta w_n=0, n\neq n_0$, one gets
\bea
\delta F=\frac{\epsilon}{2\pi\text{i}}\frac{\alpha_{n_0}}{(n_0+1)}
 \left(S_z\right)^{n_0+1}|_{z=z_0},
\label{Fvar1}
\eea
and, respectively,
\bea
\delta \wt S=-\frac{\epsilon}{2\pi\text{i}}\frac{\alpha_{n_0}}{(n_0+1)}
D(z)\left(S_z(z_0)\right)^{n_0+1}.
\label{Svar}
\eea
Taking superposition of elementary variations (\ref{Fvar1}),
we obtain a general variation of the form
\bea
\delta F=\epsilon f(S_z(z_0)),
\label{Fvar2}
\eea
where $f$ is an arbitrary analytic function (summation over
different points and integration over $z_0$ are also possible).

The formulae (\ref{Svar}), (\ref{Fvar1}) 
can be also derived considering the deformations of
nonlinear Beltrami equation (\ref{ddbar}).\\
\textbf{Remark 3.} Since a variation of the $\dbar$-data
$W$ transforms solution of the dKP hierarchy into
another solution, then the formula (\ref{Fvar2})
defines an infinitesimal symmetry transformation for the function $F$
satisfying equation (\ref{dKPadd}). It is possible to prove 
this statement directly starting with the formula 
\be
p(z_0)-p(z)+z\exp(-D(z)S(z_0))=0,\quad z_0\in\mathbb{C},
\quad z\in\mathbb{C}\setminus D.
\label{DE2}
\ee
Derivation of this formula can be found in, e.g., \cite{BKA}.
\subsection{Determinant form of equation for $\phi$}
Existence of symmetry transformation of the form (\ref{Fvar2})  leads us 
directly to equation for the function $\phi=S_z(z_0)$.
Indeed, let us consider a special symmetry transformation
(\ref{Fvar2}) of the form
$$
F' = F + \epsilon \exp(\Theta \phi),
$$
where $\Theta$ is an arbitrary parameter,
and substitute it to (\ref{dKPadd}).
Then we get a system of linear equations
\begin{eqnarray*}
\left\{
\begin{array}{l}
x+y+z=0,\\
(D_2D_3\phi)x+(D_1D_3\phi)y+(D_1D_2\phi)z=0,\\
(D_2\phi)(D_3\phi)x+(D_1\phi)(D_3\phi)y+(D_1\phi)(D_2\phi)z=0,
\end{array}
\right. 
\end{eqnarray*}
where we use notations $D_i=D(z_i)$ and
$$
x=(z_2-z_3)e^{D_2D_3F},\quad
y=(z_3-z_1)e^{D_3D_1F},\quad
z=(z_1-z_2)e^{D_1D_2F}.
$$
The condition that determinant of this system is equal to zero
gives the equation for the function $\phi$, ($\phi_i=D_i\phi$) 
\bea
\text{det}
\left(
\begin{array}{ccc}
1&1&1\\
\phi_2\phi_3&\phi_1\phi_3&\phi_1\phi_2\\
\phi_{23}&\phi_{13}&\phi_{12}
\end{array}
\right)
=0.
\label{KPconfdet}
\eea
Expanding the l.h.s. of this equation into powers of parameters
$z_1^{-1}$, $z_2^{-1}$, $z_3^{-1}$,
in the order $z_1^{-1}z_2^{-2}z_3^{-3}$ one gets the equation
\bea
\partial_x\left(\frac{\phi_t}{\phi_x}-\frac{3}{8}\left(\frac{\phi_y}{\phi_x}\right)^2  
\right)=\frac{3}{4}\partial_y \left(\frac{\phi_y}{\phi_x}\right),
\eea
which is the KP singular manifold equation in dispersionless limit.

It is interesting to note that equation (\ref{KPconfdet})
written in the form 
\bea
D_1\log\left(\frac{D_2\phi}{D_3\phi} \right)+
D_2\log\left(\frac{D_3\phi}{D_1\phi} \right)+
D_3\log\left(\frac{D_2\phi}{D_3\phi} \right)=0
\label{dMen}
\eea
has been derived in \cite{KS} as a naive continuous limit
of the discrete M\"obius-invariant KP equation
\cite{BK} (see equation (\ref{singmana}))
in connection with the Menelaus theorem. It is easy to check that
equation (\ref{KPconfdet}) (or, equivalently, (\ref{dMen}))
is invariant under conformal transformation of dependent variable
$\phi\rightarrow f(\phi)$, where $f$ is an analytic function.

The determinant form (\ref{KPconfdet}) gives some hint for the geometric
interpretation of this equation. Indeed, as it is known (see, e.g., \cite{Kraft}), 
the area of the
plane triangle with the coordinates of vertices given by the pairs
$(x_1,y_1)$, $(x_2,y_2)$, $(x_3,y_3)$, can be written as
\beaa
A=\frac{1}{2}\left|\text{det}\mathbf{A}\right|, \qquad
\mathbf{A}=\left(
\begin{array}{ccc}
1&1&1\\
x_1&x_2&x_3\\
y_1&y_2&y_3
\end{array}
\right).
\eeaa
Thus equation (\ref{KPconfdet}) means that area (may be complex) 
of corresponding triangle vanish, that is, obviously, the only 
possibility for the geometry admitting arbitrary transformation
$\phi\rightarrow f(\phi)$.

More general generating equation for the gauge-invariant
function $\mathcal{S}=S(z_1)-S(z_0)$
\bea
\sum \epsilon_{ijk}D_j\log(\exp(D_i \mathcal{S})-1)=0
\label{dKPsm}
\eea
derived in \cite{KM2}
(from which equation (\ref{dMen}) can be easily obtained by the limit
$z_1\rightarrow z_0$)
can be also written in the determinant form,
\bea
\text{det}
\left(
\begin{array}{ccc}
1&1&1\\
(e^{\mathcal{S}_2}-1)(e^{\mathcal{S}_3}-1)& 
(e^{\mathcal{S}_3}-1)(e^{\mathcal{S}_1}-1)&
(e^{\mathcal{S}_1}-1)(e^{\mathcal{S}_2}-1)\\
\mathcal{S}_{23} & \mathcal{S}_{13}& \mathcal{S}_{12}
\end{array}
\right)
=0,
\nn
\eea
where the subscript $i$ refers to the derivative $D_i$. 
We believe it is possible to obtain this equation in a manner
similar to the derivation of (\ref{KPconfdet}).
\subsection{Determinant form of discrete SKP equation}
The basic idea of derivation of the determinant form
(\ref{KPconfdet}) is applicable to the dispersionfull case too
(in fact, originally we used it in the dispersionfull case first).
This idea gives another way to obtain discrete Swartzian
KP equation, which was introduced in \cite{BK}.
Instead of equation (\ref{dKPadd}), we start from the well-known
addition formula for the KP $\tau$-function
\bea
c_1(T_1 \tau)(T_2T_3\tau)+
c_2(T_2 \tau)(T_1T_3\tau)+
c_3(T_3 \tau)(T_1T_2\tau)=0,
\label{tauadd}
\eea
where $T_i$ denotes a Sato shift, $c_i$ are certain coefficients.
\begin{theorem}
If for some function $\Phi$ the function
\bea
\widetilde\tau=\tau(1+\Theta\Phi),
\label{tautrans}
\eea
satisfies equation (\ref{tauadd}) for arbitrary $\Theta$
(i.e., formula (\ref{tautrans}) defines a
B\"acklund transformation for the $\tau$-function),
then the function $\Phi$ satisfies the equation
\bea
\det
\left(
\begin{array}{ccc}
1&1&1\\
(T_1\Phi+T_2T_3\Phi)&(T_2\Phi+T_1T_3\Phi)&(T_3\Phi+T_1T_2\Phi)\\
(T_1\Phi)(T_2T_3\Phi)&(T_2\Phi)(T_1T_3\Phi)&(T_3\Phi)(T_1T_2\Phi)
\end{array}
\right)
=0.
\label{dSKPdet}
\eea
\end{theorem}
\textbf{Proof} 
Substituting (\ref{tautrans}) into (\ref{tauadd}),
we obtain a system of linear equations
\begin{eqnarray*}
\left\{
\begin{array}{l}
x+y+z=0,\\
(T_1\Phi+T_2T_3\Phi)x+(T_2\Phi+T_1T_3\Phi)y+(T_3\Phi+T_1T_2\Phi)z=0,\\
(T_1\Phi)(T_2T_3\Phi)x+(T_2\Phi)(T_1T_3\Phi)y+(T_3\Phi)(T_1T_2\Phi)z=0,
\end{array}
\right. 
\end{eqnarray*}
where
$$
x=c_1(T_1 \tau)(T_2T_3\tau),\quad
y=c_2(T_2 \tau)(T_1T_3\tau),\quad
z=c_3(T_3 \tau)(T_1T_2\tau).
$$
Then, from the requirement that determinant of this linear system
should be equal to zero, we get equation (\ref{dSKPdet}), QED.

It is easy to check that equation (\ref{dSKPdet}) coincides
with discrete SKP equation, introduced in \cite{BK}
in multiplicative form
\bea
(T_2\Delta_1 \Phi)(T_3\Delta_2 \Phi)(T_1\Delta_3 \Phi)=
(T_2\Delta_3 \Phi)(T_3\Delta_1 \Phi)(T_1\Delta_2 \Phi),
\label{singmana}
\eea
where $\Delta_i=T_i-1$ is a difference operator.
\section{Dispersionless 2DTL hierarchy}
Most of the results presented above for dispesionless KP hierarchy
are valid after some modification for dispersionless 2DTL 
hierarchy. First we will outline the basic notations, following
the work \cite{BKA}, and then we will discuss how the main formulae 
are modified.

For the dispersionless 2DTL hierarchy the $\dbar$-data are localized
on the domain $G$
which is  an annulus
$a<|z|<b$, where $a,b$ ($a,b\in\mathbb{R}$, $a,b>0$; $b>a$) 
are arbitrary (instead of the unit disc in KP case).
To set the quasi-classical $\dbar$-problem (\ref{ddbar}) correctly,
in general we do not need to require analyticity of the function
$S_0$ in $G$, it is enough to have analyticity of 
its derivative ${S_0}_z$.
A generic function $S_0$ 
with ${S_0}_z$ analytic in $G$ can be represented
as
$$
S_0(t,\mathbf{x},\mathbf{y})=t\log z +\sum_{n=1}^{\infty}z^n x_n+
\sum_{n=1}^{\infty}z^{-n}y_n,
$$
where $t,x_n,y_n$ are free parameters \cite{KM2}.
We assume that $\wt S(z)\sim \sum_{n=1}^\infty \frac{S_n}{z^n}$
as $z\rightarrow\infty$ and denote $\wt S(0)=\phi$,
$G_+=\{z,|z|>b\}$, $G_-=\{z,|z|<a\}$.
Relations, characterizing the $\tau$-function $F$, are 
$\phi=DF$, $\wt S(z_1)=-D_+(z_1)F$ $(z_1\in G_+)$,
$\wt S(z_2)=\phi-D_-(z_2)F$ $(z_2\in G_-)$, where
$D_+(z)=\sum_{n=1}^\infty\frac{1}{n}\frac{1}{z^n}\frac{\partial}{\partial x_n}$,
$D_-(z)=\sum_{n=1}^\infty\frac{1}{n}{z^n}\frac{\partial}{\partial y_n}$,
$D=\frac{\partial}{\partial t}$.
\subsection{$\tau$-function for dispersionless 2DTL hierarchy}
Boundary problem for nonlinear Beltrami equation (\ref{ddbar})
in this case is formulated on the boundary of the annulus $G$,
and integration in the formula (\ref{TAU}) goes over the annulus.
\begin{theorem}
The function
\bea
F(t,\mathbf{x},\mathbf{y})
=\frac{-1}{2\pi\text{i}}\iint_{G} 
\left(\frac{1}{2}\wt S_{\bar z}(t,\mathbf{x},\mathbf{y}) 
\wt S_z(t,\mathbf{x},\mathbf{y})-
W_{-1}(z,\bar z,S_z(t,\mathbf{x},\mathbf{y}))\right)dz\wedge d\bar z
\nn
\eea
is a $\tau$-function of dispersionless 2DTL hierarchy.
\end{theorem}
The proof is completely analogous to the KP case.

The formula for the $\tau$-function
may be also considered in a more general
context, for arbitrary domain $G$. In this case it defines $F$
as a functional on the space of functions $S_0(z)$, having
the derivative ${S_0}_z$ analytic in $G$. The form of corresponding hierarchy depends on parameterization of this space in terms of
`times'.

\subsection{Symmetries and singular manifold equations}
Variations of the function $F$, preserving the hierarchy,
have the same form (\ref{Fvar2}) as in KP case. 

Dispersionless 2DTL equations can be obtained from KP case
(equations (\ref{dKPsm}), (\ref{dMen}), (\ref{KPconfdet}) )
by the transformations
$D_1\rightarrow D_+$, $D_2\rightarrow D_- + \partial_t$,
$D_3\rightarrow \partial_t$
or, equivalently,
$
D_1\rightarrow D_-$, $D_2\rightarrow D_+ - \partial_t$,
$D_3\rightarrow -\partial_t
$,
connected by the the symmetry
$
D_+\rightarrow D_-$, $D_-\rightarrow D_+$,
$\partial_t\rightarrow -\partial_t$.

Equation (\ref{dKPsm}) for $\mathcal{S}$ takes the form
\bea
&&
\left(e^{D_+\mathcal{S}}(e^{-\partial_t \mathcal{S}}-1)+ 
e^{D_-\mathcal{S}}(e^{\partial_t \mathcal{S}}-1)\right)
D_+D_- \mathcal{S}
\nn\\&&\qquad\qquad
-(e^{D_+\mathcal{S}}-1)(e^{D_-\mathcal{S}}-1)
\partial_t(D_- -D_+ +\partial_t)\mathcal{S}=0.
\label{TodaS}
\eea
In the order $z_+(z_-)^{-1}$ of expansion of equation (\ref{TodaS})
one gets
\bea
\mathcal{S}_{xy}-
\mathcal{S}_x\mathcal{S}_y\frac{\mathcal{S}_{tt}}{(e^{\mathcal{S}_t}-1)
(1-e^{-\mathcal{S}_t})}=0,
\eea
that is the dispersionless 2DTL singular manifold equation.

Generating equation for conformally invariant case reads
\bea
(\partial_t\phi)(D_+D_-\phi)(\partial_t+D_- - D_+)\phi=
(D_+\phi)(D_-\phi)\partial_t(\partial_t+D_- - D_+)\phi.
\eea
The order of expansion $z_+(z_-)^{-1}$
gives the conformally invariant dispersionless 2DTL equation
\bea
\frac{\phi_{xy}}{\phi_x\phi_y}=\frac{\phi_{tt}}{\phi_t\phi_t}.
\eea
\subsection*{Acknowledgments}
LVB was supported in part by RFBR grant
01-01-00929 and President of Russia grant 1716-2003; BGK was supported in part
by the grant COFIN 2000 `Sintesi'.

\end{document}